\documentstyle[emulateapj,epsf]{article}

\begin{document}

\submitted{To Appear in ApJ letters}

\title{\Large Dark matter scaling relations}

\author{P.~Salucci} 

\affil{ SISSA - International School for Advanced Studies, via Beirut 2-4,
        I-34013 Trieste, Italy }
\author{ A.~Burkert}

\affil{        Max-Planck-Institut f\"ur Astronomie, K\"onigstuhl 17, D-69117
        Heidelberg, Germany}

\begin{abstract}
We investigate the structure of dark matter halos by means of the kinematics of 
 a very large sample of spiral galaxies of all  luminosities.

The observed rotation curves
show  an  universal profile which is the sum of 
an exponential thin disk term and a spherical halo term with a flat  density core. We find that
 the Burkert  profile proposed to  describe  the dark matter halo density distribution
of dwarf galaxies also provides an excellent mass model for the dark halos
around  disk systems up to 100 times more massive.
Moreover, we find that spiral  dark matter core densities
$\rho_0$ and core radii $r_0$ lie in  the same
scaling relation $\rho_0
 = 4.5 \times 10^{-2} (r _0/kpc)^{-2/3} M_{\odot}
pc^{-3} $ of dwarf galaxies with core radii up to  10 times smaller.

At  the highest masses $\rho_0$   decreases
 with $r_0$ faster than the   $-{2\over {3}}$ power law implying     a  lack of
 objects 
 with disk masses  $> 10^{11}M_\odot$ and  central densities 
$> 1.5 \times 10^{-2}(r_0/kpc)^{-3}~M_{\odot}
pc^{-3}$  that can be explained by the existence of  a
{\it maximum} mass  of $\approx 2 \times 10^{12} M_{\odot}$ for an halo   hosting 
a spiral galaxy.

\end{abstract}

\begin{keywords} {Galaxies: spiral, kinematics and dynamics Cosmology: dark matter}
\end{keywords}

 \section{Introduction}

It is now well established that spiral galaxies have universal rotation 
curves (URC) that can be characterized by one single free parameter, the
luminosity (e.g. Rubin et al. 1980; Persic  \& Salucci, 1991, 
Persic, Salucci \& Stel 1996 (PSS)). For instance,  low-luminosity spirals
show  ever-rising rotation curves  (RC) out to 
 the optical radius \footnote{ $R_{opt} = 3.2 R_d$, with $R_d$ the exponential
disk length-scale}, while,  in the same 
region, the RC of   high-luminosity spirals 
 are flat or even decreasing.

It has been demonstrated  by Persic \& Salucci (1988, 1990)
and Broeils (1992) that, as the  galaxy luminosity decreases,  the light is progressively  unable
to trace the radial distribution of the gravitating  matter (see also Salucci, 1997).
This  discrepancy  is, in general,
interpreted as the  signature of an  invisible mass
component (Rubin et.al. 1980; Bosma 1981). 
As pointed out  by PSS the universality of
the rotation curves, in combination with the invariant  distribution of the luminous 
matter, implies  an universal dark matter distribution
with luminosity-dependent scaling properties.
 
On  the theoretical side,  recent high-resolution cosmological N-body simulations have 
 shown that cold dark matter halos achieve a specific  
equilibrium density profile (Navarro, Frenk
\& White 1996, NFW; Cole \& Lacey 1997; Fukushige \& Makino 1997;
Moore et al. 1998; Kravtsov et al. 1998). This can be characterized
by one free parameter, e.g.  $M_{200}$, the dark mass
enclosed within    the radius  inside which
the average over-density is 200 times the critical density of the Universe.
In the innermost regions  the dark matter profiles show some 
scatter around an average  profile which  is characterized by a power-law cusp
 $\rho \sim r^{-\gamma} $,  with $\gamma =1-1.5$ (NFW, Moore et al. 1998, 
Bullock et al., 1999).

Until recently,   due  to both  the limited number 
of suitable rotation curves and  a poor knowledge   of 
 the exact  amount of
luminous matter present in the innermost  regions of spirals,  
it has been  difficult
to investigate    the  internal
structure of dark matter halos.
The situation is  more favorable  for (low surface
brightness) dwarf galaxies which are   strongly dark matter dominated even
at small radii.  The kinematics  of these systems  shows  an  universality
of the dark halo density profiles, but, it results in 
disagreement with that predicted by CDM, in particular because of    
the existence of  dark halo density  cores.
  (Moore 1994; Burkert 1995). 
The origin of these features  is not yet   
understood (see e.g. Navarro, Eke \& Frenk 1996; Burkert \& Silk
1997, Gelato \& Sommer-Larsen 1999),  but it is likely  that it  involves 
more physics  than  a simple  hierarchical assembly of cold structures.
To cope with this observational  evidence, Burkert (1995) proposed an empirical profile
that successfully  fitted  the halo  rotation curves of
four dark matter dominated dwarf galaxies 
\begin{equation}
\rho_b(r) = \frac{\rho_0 r^3_0}{(r+r_0)(r^2+r_0^2)}
\end{equation}
\noindent
 where $\rho_0$ and $r_0$ are free parameters which represent
the central dark matter density and the scale radius. 
This sample has been extended,  more recently,  to 17 dwarf  irregular 
and low surface brightness galaxies (Kravtsov et al. 1998, see however
van den Bosch et al. 1999) which all  are found to  confirm  equation (1).
Adopting spherical symmetry, the mass distribution of the Burkert halos
is given by
\begin{equation}
M_b(r) = 4~M_0 \Big\{ ln \Big( 1 + \frac{r}{r_0} \Big) -tg^{-1} \Big(   \frac{r}{r_0} \Big) +{1\over {2}} ln 
\Big[ 1 +\Big(\frac{r}{r_0} \Big)^2 \Big] \Big\}
\end{equation}
\noindent
with $M_0$,  the dark 
mass within the core given by:
\begin{equation}
M_0 = 1.6 \rho_0 r_0^3
\end{equation}
\noindent
The halo contribution to the circular  velocity is then:  
\begin{equation}
V^2_b(r)=GM_b(r)/r.
\end{equation}
Although the dark matter core parameters $r_0$, $\rho_0$ and $M_0$ are
in principle independent,   the observations reveal a clear  correlation
(Burkert 1995):
\begin{equation}
M_0  =  4.3\times 10^7 \left( \frac{r_0}{kpc} \right)^{7/3} M_{\odot} 
\end{equation}
\noindent
 which indicates that dark  halos represent
a 1-parameter family which is completely specified,  e. g. by the core mass.

\vbox{ \vskip 1.truecm
\centerline{\epsfxsize=4.8truecm
\epsfbox{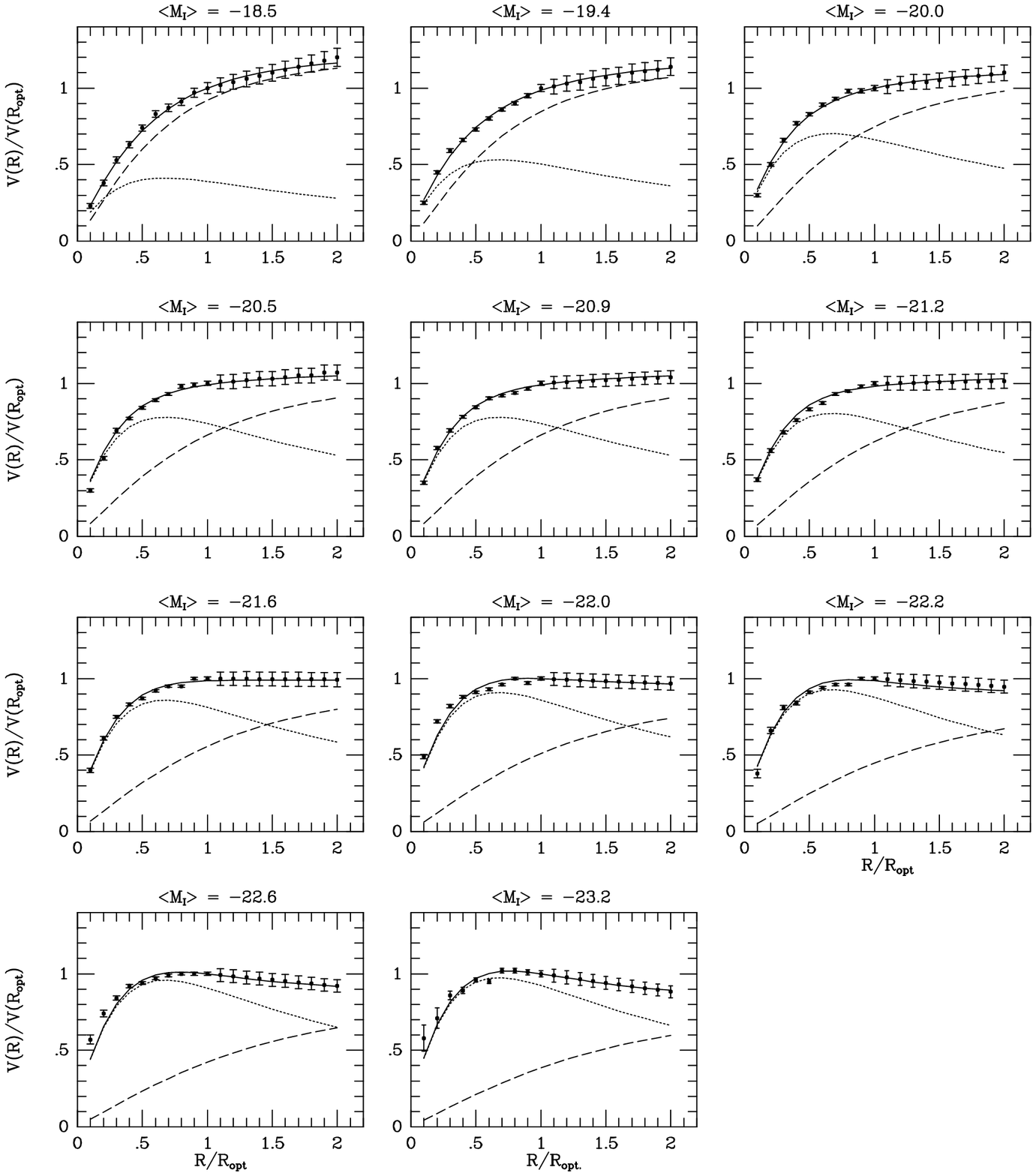}}
\vskip 0.00truecm \figcaption[]{ Synthetic rotation curves (filled circles with error bars)
and  URC   (solid line) with  its  separate dark/luminous  
contributions (dotted line: disk;  dashed line: halo.) See PSS for details.  } \vskip 0.7truecm }

 The analysis of a   recently published  large sample of  RC's (Persic \& Salucci, 1995)  
has  provided a suitable framework    to 
investigate the dark halo density  distribution in spirals.
 The starting  points of this study  are: {\it a)} the mass in spirals is distributed 
according to   the  Inner Baryon Dominance (IBD) regime: there is    
a characteristic transition  radius
$R_{IBD} \simeq 2 R_d  \Big({V_{opt}\over{220 km/s}}\Big)^{1.2}$
 for which,  at  $r\leq R_{IBD}$,    the luminous matter  totally
 accounts for the mass distribution,   while, for   $r> R_{IBD}$,  
the DM {\it rapidly}
 becomes the  dominant dynamical component (Salucci and Persic, 1999a,b;  Salucci et al,
2000;  Ratnam and Salucci, 2000;  Borriello and Salucci, 2000). 
Then,  although the  dark halo might  extend  down to the galaxy 
center, it is only  for   $r>R_{IBD}$ that it  gives  a non-negligible contribution to the
circular velocity. {\it b)}  The dark matter  is
distributed in a   different way with respect to any of  the various    baryonic
components  (PSS, Corbelli and Salucci, 2000), and {\it c)}  The HI  contribution to the
circular velocity, for $r< R_{opt}$, is negligible (e.g.  Rhee, 1996 ;
Verheijen, 1997).

The main aim of this  letter is to expand   the above  results   to  derive 
the luminosity-averaged density profiles  of the  dark halos   and   to 
 relate   them  with the   Burkert profiles. 
Section 2 presents the  analysis of  a homogeneous sample of about 1100 
rotation curves in which the dark halo contribution to the circular
velocity is first derived and then  matched  
to the Burkert halo  mass models. 
In section 3 we discuss  the results. We take $H_0=75 km/s/Mpc$
and $\Omega_0=0.3$, however no result depends on these choices.

\section{DM halo profiles in Spirals}

PSS (see also Rubin et al. 1982)
 have   derived,  from   $15000$
velocity measurements of   1000  
rotation curves (RC),    $V_{syn}({r\over{R_{opt}}}, {L_I\over{L_*}})$,
the synthetic rotation  velocities  of spirals   sorted by 
luminosity  (Figure 1,  with $L_I$ the
I-band luminosity and  $L_I/L_*=10^{-(M_I+21.9)/5})$.  Remarkably,    {\it individual}  RC's   have 
a very small variance with respect to the corresponding  synthetic curves
 (PSS, Rhee 1996,
Rhee \& van Albada 1996, Roscoe 1999, Swaters, 1999): 
spirals sweep a very narrow locus in the 
RC-profile/amplitude/luminosity space. On the other hand, the galaxy kinematical
properties do significantly change with luminosity (e.g. PSS), so it is  
natural  
to relate    the  mass distribution with this quantity.  
The whole set of  synthetic RC's  has been  
modeled  with  the Universal Rotation Curve (URC),  $V_{URC}(r/R_{opt}, L_I/L_*)$  
which includes:  {\it (a)} an
exponential thin disk term:
 \begin{equation}
V^2_{d,URC}(x)=1.28~\beta V^2_{opt}~ x^2~(I_0K_0-I_1K_1)|_{1.6x}
\end{equation}
and
{\it (b)} a spherical halo term:
\begin{equation}
V_{h,URC}^2(x)= V^2_{opt}  ~(1-\beta) ~(1+a^2) {x^2 \over{(x^2+a^2})}\,,
\end{equation}
with $x \equiv r/R_{opt}$, 
$\beta
\equiv (V_{d,URC}(R_{opt})/V_{opt})^2$, $V_{opt} \equiv V(R_{opt})$ and  
$a$ the halo core radius in units of $R_{opt}$. At high luminosities 
the contribution from a bulge component has been also considered (Salucci and Persic, 1999b).

\vbox{ \vskip -3.3truecm
\centerline{\epsfxsize=8.3truecm
\epsfbox{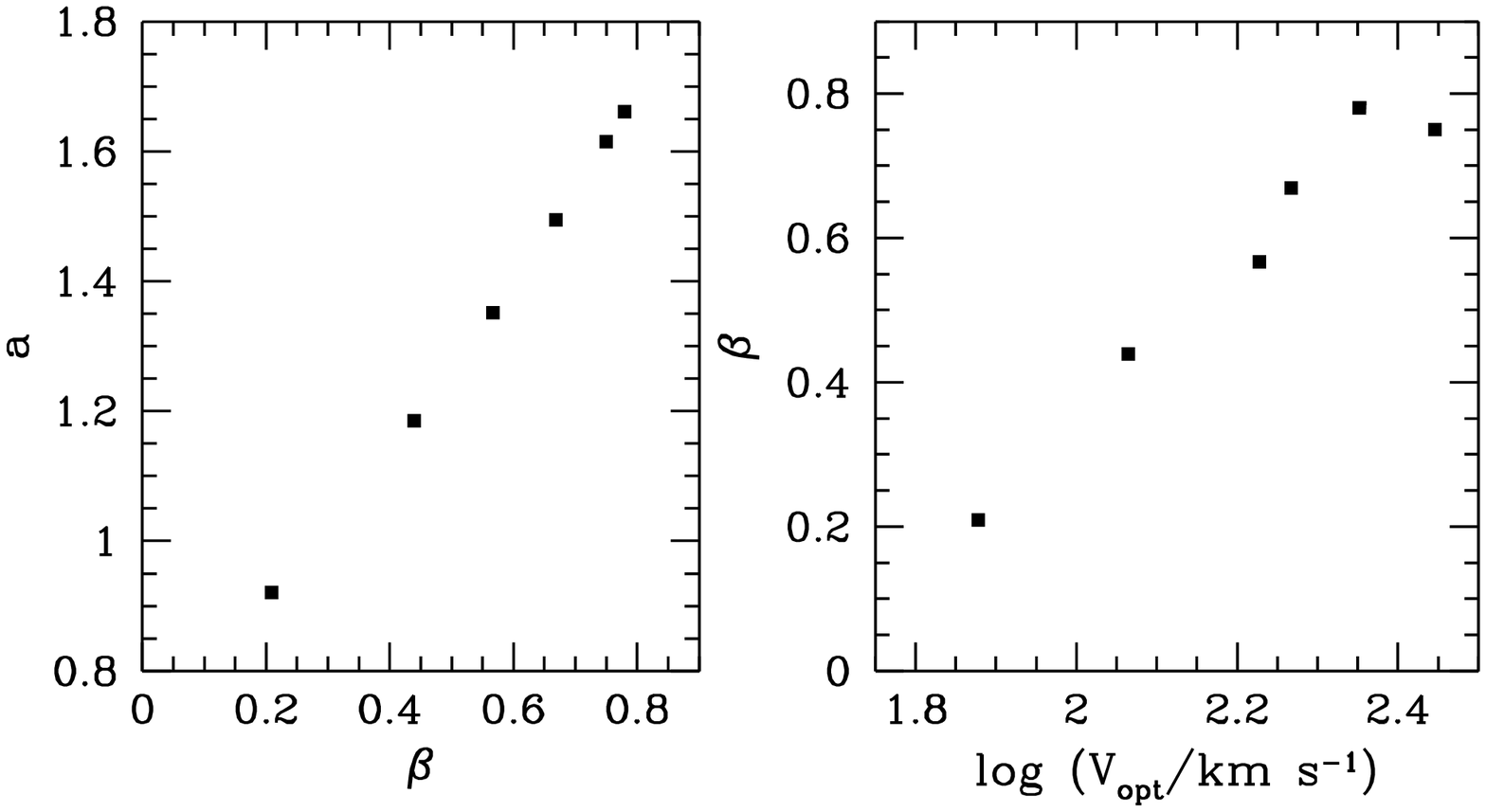}}
\vskip - 0.3truecm \figcaption[]{ $a$ {\it vs} $\beta$  and 
$\beta$ {\it vs} $V_{opt}$. } \vskip .2truecm }
Let us stress  that the halo  velocity  term  of  eq. (7)  does not bias     
the mass model:    it can   account for  
maximum-disk, solid-body,  no-halo, all-halo,
 CDM   and
 core-less halo mass  models. In practice, the  values of the  
free parameters $a$ and $\beta$  that  result from  fitting the URC
 \begin{equation}
V_{URC}^2(x)=V^2_{h,URC}(x,\beta,a)+V^2_{d, URC}(x,\beta)
\end{equation}
to the  synthetic  curves $V_{syn}$    
select the actual   model. 
Adopting $\beta ~\simeq  0.72 +~ 0.44\, log  ({L_I \over L_*})$ 
and $a \simeq  1.5  ( {L_I \over L_*} )^{1/5}$
(see PSS)  or,  equivalently the corresponding  
\begin{equation}
a =a(\beta) \ \ \ \ \ \ \ \ \  \ \\  \ \ \ \ \ \ \ \beta=\beta (log \ V_{opt}) 
\end{equation}
\noindent 
that we have plotted  in  Fig.  (2), 
the URC mass models 
reproduce the synthetic curves  $V_{syn}(r)$  within  their r.m.s.  (see Fig. (1)). More in  detail,  at any   
luminosity and radius, $|V_{URC}-V_{syn}| <  
 2\%$ and the $1\sigma$ fitting  uncertainties  on
$a$ and $\beta$  are about 
20\%  (PSS). 

We then  compare  the dark halo velocities obtained  with  eq. (7) and (9),
with   the Burkert velocities   $V_b(r)$ of   eqs.  (2)-(4). We 
leave    $\rho_0$ and $r_0$  as free parameters, i. e.  we do not impose 
the relationship of
eq. (5).   The results are 
shown in Fig (3): at any luminosity, out to the outermost  radii ($\sim 6 R_d$),
$V_b(r)$ is  indistinguishable from $V_{h,URC}(r)$.   More specifically,  by
setting   $V_{h, URC}(r)\equiv V_{b}(r)$,  we are able to reproduce     
  the  synthetic rotation curves $V_{syn}(r)$ at the level of  their  r.m.s. 
 For $r>>6 R_D$,  i.e. beyond the region described by the 
URC, the two velocity  profiles    progressively differ.  

\vbox{ \vskip 0.5truecm
\centerline{\epsfxsize=10.5truecm
\epsfbox{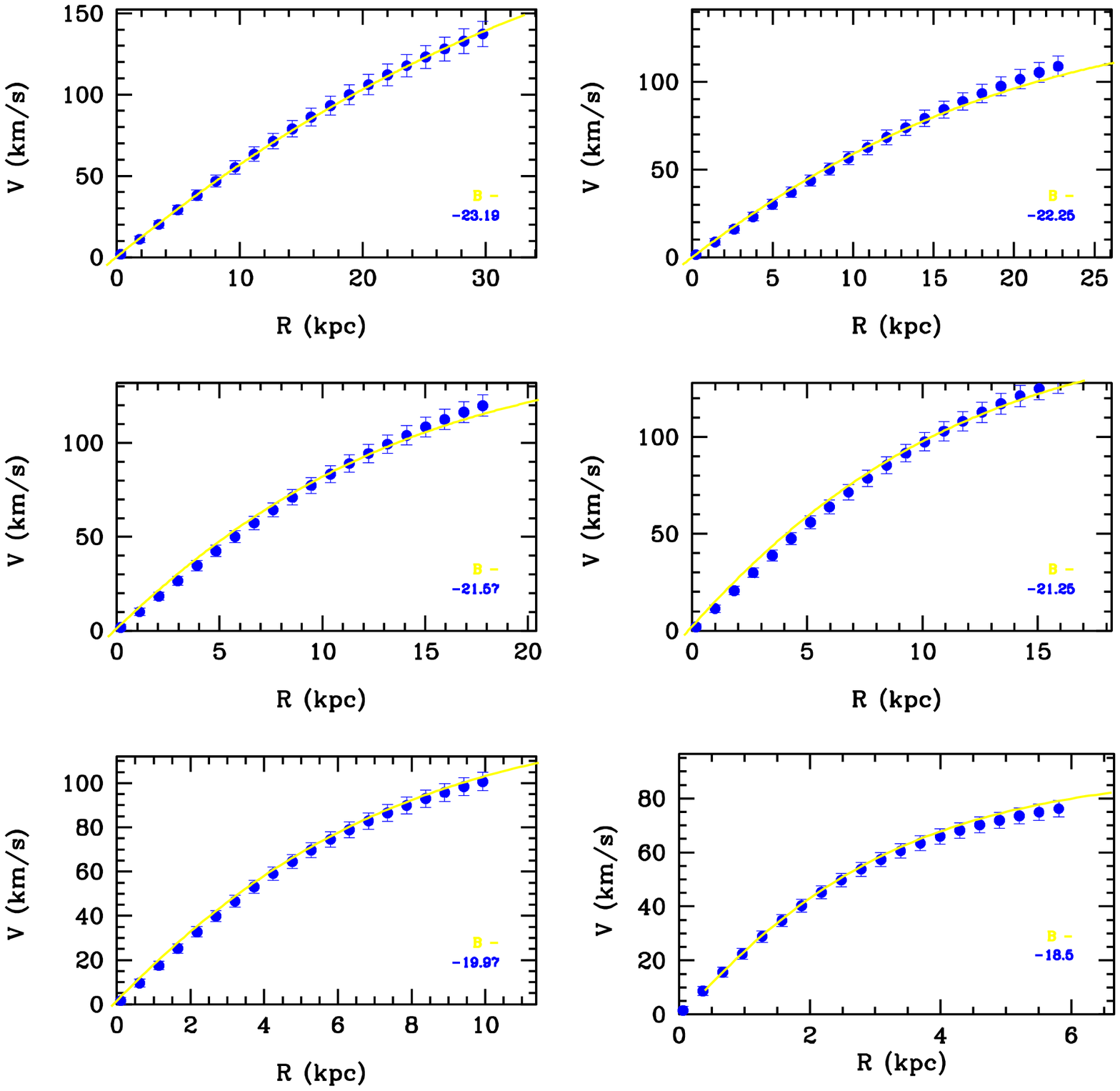}}
\vskip 0.00truecm \figcaption[]{ URC-halo  rotation curves  
({\it filled circles} with error
bars)
and the Burkert model ({\it solid line}). 
The bin magnitudes are also indicated.
}
\vskip 1.5truecm 
}  

The values obtained for $r_0$ and $\rho_0$ for the URC  
agrees with the extrapolation at high masses
of the  scaling law  $\rho \propto r_0^{-2/3} $ (Burkert, 1995)  
established for objects with core radii $r_0$ ten times
smaller (see  Fig 4). Let us notice  that the core radii are very 
large: $r_0>>R_d$  so that an 
ever-rising halo RC cannot be excluded by  the  data.
Moreover, the  disk-mass vs. central halo  density
 relationship $\rho_0 \propto M_d^{-1/3}$, found  in   
dwarf galaxies (Burkert,1995), 
where the  densest  halos harbor the least massive disks, 
holds  also  for   disk systems  of stellar  mass up to  
$10^{11} M_\odot$
(see Fig 4).

The above  relationship  show a curvature at the highest  masses/lowest densities
that can be related  to  the existence of 
an upper limit in  the  dark halo mass $M_{200}$ \footnote{The virial
halo mass is given by; $M_{200}\equiv 200 \times 4\pi/3 \rho_c R_{200}^3 \Omega_0 (1+z^3) g(z) $
with $z$ the formation redshift, $R_{200} $ the virial radius, for   $g(z)$ 
see e.g.  Bullock
et al., (1999);  the critical 
density is defined as:  $\rho_c\equiv 3/(8  \pi) G^{-1} H_0^2$.} which  is
evident by  the sudden  decline  of  the  baryonic {\it mass} function of 
disk galaxies at $M_{d}^{max}= 2\times  10^{11}M_\odot $ 
(Salucci and Persic, 1999a), that implies a maximum halo mass of 

\vspace{-3.0cm}
\begin{equation}
 M_{200}^{max}  \sim \Omega_0/\Omega_{b} ~ M_{d}^{max}
\end{equation}
\vspace{-3.0cm}

where $\Omega_0$ and $\Omega_{b}\simeq 0.03$ (e.g. Burles and Tytler, 1998)
 are  the  matter  and baryonic
densities of the Universe in units
of  critical density. From the definition of $M_{200}$,  by means
of  eq. (2) and (3), we can write  $M_{200}$  in terms of  the "observable" quantity  $M_{0}$:
 $M_{200} = \eta M_0$.
For $(\Omega_0, z)=(0.3,3)$, $\eta \simeq 12$;  
notice that there is a   mild  dependences of  $\eta$ 
on  $z$ and $\Omega_0$ which is   irrelevant for the present study. 
Combining  
eq. (3) and (10) we obtain an upper limit  for the central density,  $\rho_0 <1\times 10^{-20}
(r_0/kpc)^{-3} g /cm^{3}$,  which  implies  a  lack of objects with $\rho_0> 4\times
10^{-25}g/cm^{3}$ and $r_0>30 kpc$,  as is evident in Figure 3. 
Turning the argument around, the  deficit   of
objects with $M_d \sim M_d^{max} $ and   $\rho_0> 4\times 10^{-25}~g/cm^{3}$, suggests  
 that, at this mass scale, 
the total-to- baryonic mass  fraction may approache the cosmological value  $\Omega/\Omega_{b}\simeq 10$.  
\vspace{-1.0cm}
\section{Discussion}

Out to two optical radii,  the  Burkert density profile  reproduces,  
for the whole  spiral  luminosity sequence, the DM halos mass distribution.
This density   profile, though  at very large radii  coincides
with the NFW profile, approaches a constant, finite density value at the center,  in a way
consistent with an isothermal distribution. This is in contradiction to 
cosmological 
models (e.g. Fukushige and Makino 1997) which predict that the velocity dispersion $\sigma$ of the
dark matter particles decreases towards the center to reach 
$\sigma \rightarrow 0$ for $r \rightarrow 0$.
After the result of this study, the dark halo inner regions,  
therefore, cannot be considered as kinematically cold structures 
but rather as     
"warm"  regions with size $ r_0 \propto  \rho_0^{-1.5}$.  
The halo core sizes  are  very large:  $r_0 \sim 4-7 R_d$.
Then, the boundary of the core region is  well beyond  the region   
where the stars are located and,
as in  Corbelli and Salucci (2000), even at  the  outermost observed  radius
there is not  the  slightest  evidence that dark  halos  converge
to a    $\rho \sim r^{-2}$ (or a steeper) regime.  

\vbox{ \vskip 0.5truecm
\centerline{\epsfxsize=7.9truecm
\epsfbox{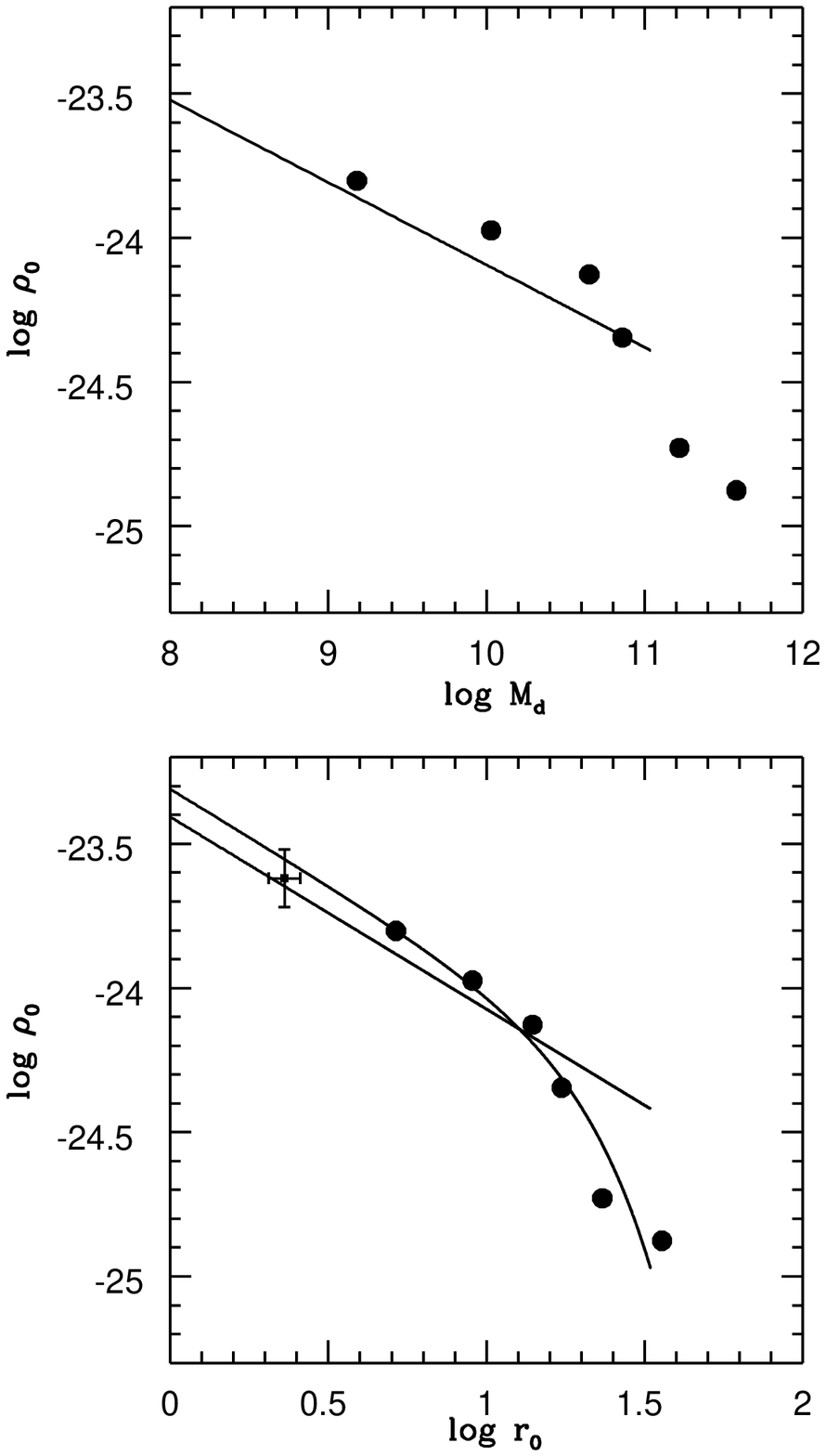}}
\vskip 0.00truecm \figcaption[]{   {\it top} 
Disk mass (in solar units) {\it vs}   central halo density  $\rho_0$ 
(in $g/cm^3$) for   normal spirals ({\it filled circles}).  
The straight line is 
the extrapolation  to high luminosities  of the   relation of  dwarfs.  
  {\it
bottom }  Central density    {\it vs}  core radii  (in kpc)  for  normal 
spirals ({\it filled circles}),   compared with  the extrapolation of the 
relationship of  dwarfs ({\it dotted line},  
the point with errorbar represents a typical  object of Burkert,  1995). 
The solid line is the   eye-ball fit;
$\rho_0=5\times 10^{-24}r_0^{-2/3}e^{-(r_0/27)^2} g/cm^3$. 
The effect of a limiting
halo mass is also  shown ({\it dashed line}).}
 \vskip 0.5truecm
 }
We find  that the dark  halos  around  spirals  are 
essentially an one-parameter family. It is relevant that the order 
parameter  (the central density or the core radius) correlates   with the
luminous  mass  (see Fig 4). We do however not know how it is
related to the global structural properties of the dark halo, like the
virial radius or the virial mass. In fact, 
the RC out to $6R_D$ is completely determined by the core parameters, 
i.e. the central core density
and the core radius,  both of which are not defined  in the CDM scenario.

The location of spiral galaxies in the parameter space of virial
mass, halo  central density and baryonic mass is determined by   
different processes that occur on different scales and at
different red-shifts. Yet, this  3D space
degenerates into  a single  curve (see Figure 4,remind that: 
 $\rho_0 = {\pi\over {24}} ~ M_{200}/r_0^{3}$ and remind that: 
 $M_d=G^{-1}\beta V^2_{opt}R_{opt})$ which describes   
the  dark-luminous coupling.

Let us  discuss the limitations   of the present results.
First,  here we have considered the luminosity
dependence of the dark halo structure.  Although  this   is  probably   
the most relevant one, other dependences (Hubble type and surface
brightness) should also be investigated.  Moreover, the existence of 
 a (weak) cosmic variance   in   the halo 
structural properties cannot be excluded  until we analyze individual 
objects (Salucci, 2000, Borriello and Salucci, 2000).  
Secondly, we have derived 
the profile of DM halos  out to  about six disk-scale lengths, i.e.
out to a distance much smaller than the
virial radius.  To assess the  {\it global}  validity of the  proposed  
mass model   
data at larger radii are obviously required.


\vfill\eject

\end{document}